\documentclass[twocolumn,english,aps,superscriptaddress,twocolumn,twoside,floatfix,prl]{revtex4-2}
\usepackage[T1]{fontenc}
\usepackage[latin9]{inputenc}
\usepackage{xcolor}
\usepackage{amsmath}
\usepackage{stackrel}

\makeatletter

\newcommand*\LyXZeroWidthSpace{\hspace{0pt}}

\usepackage{tikz}
\usepackage[colorlinks,linkcolor=blue]{hyperref}
\allowdisplaybreaks

\makeatother

\usepackage{babel}
\begin{document}
\preprint{APS/123-QED}

\title{Revisiting Schwarzschild black hole singularity through string theory}

\author{Houwen Wu}  
\email{hw598@damtp.cam.ac.uk}
\affiliation{College of Physics, Sichuan University, Chengdu, 610065, China}
\affiliation{DAMTP, Centre for Mathematical Sciences, University of Cambridge, Cambridge, CB3 0WA, UK}

\author{Zihan Yan}  
\email{zy286@cam.ac.uk}
\affiliation{DAMTP, Centre for Mathematical Sciences, University of Cambridge, Cambridge, CB3 0WA, UK}

\author{Shuxuan Ying}  
\email{ysxuan@cqu.edu.cn} 
\affiliation{Department of Physics, Chongqing University, Chongqing, 401331, China}

\date{\today}

\begin{abstract}
In this letter, we derive the singular condition for black holes and demonstrate the  potential resolution of the Schwarzschild black hole singularity in general relativity using non-perturbative $\alpha^{\prime}$ corrections of string theory. This work is motivated by the Belinskii, Khalatnikov and Lifshitz (BKL) proposal, which suggests that the structure of the black hole interior in vacuum Einstein's equations can be transformed into the Kasner universe near the singularity. This transformation allows for the description of the black hole interior using the $O\left(d,d\right)$ invariant anisotropic Hohm-Zwiebach action, which includes all orders of $\alpha^{\prime}$ corrections.

\end{abstract}

\maketitle

Black holes are the most important objects in our universe, as understanding
their internal structure and microstate could reveal the fundamental
nature of physics. However, gravitational collapse inevitably leads
to a curvature singularity at the center of a black hole, a feature
that remains unavoidable in the framework of Einstein's gravity \cite{Penrose:1964wq,Hawking:1965mf}.
This suggests that general relativity is incomplete and breaks down
in the vicinity of this singularity. String theory, a candidate theory
of quantum gravity, is expected to modify the general relativity by
including higher-derivative corrections and resolving this longstanding
issue. However, to date, it has only provided insights into two-dimensional
string black holes \cite{Witten:1991yr,Dijkgraaf:1991ba}, leaving
unresolved the singularity problem of the four-dimensional black holes
of Einstein's gravity. This unbridgeable gap arises from the unknown
non-perturbative aspects of string theory when approaching the curvature
singularity.

In this letter, we aim to investigate whether the singularity of the
Schwarzschild and rotating, accreting black holes can be resolved
by higher-derivative $\alpha^{\prime}$ corrections of string theory.
This work is motivated by the recent progress in classifying all orders
of $\alpha^{\prime}$ corrections for specific backgrounds by Hohm
and Zwiebach \cite{Hohm:2015doa,Hohm:2019ccp,Hohm:2019jgu}. In previous
work, Sen has demonstrated that when the classical fields, including
the metric, Kalb-Ramond field and dilaton, are independent of $m$
coordinates, the field theory exhibits an $O\left(m,m\right)$ symmetry
to all orders in $\alpha^{\prime}$ \cite{Sen:1991zi,Sen:1991cn}.
This result has been verified for a few orders in $\alpha^{\prime}$,
indicating that the standard form of $O\left(d,d\right)$ transformations
can be preserved through suitable field redefinitions, even in the
presence of $\alpha^{\prime}$ corrections \cite{Veneziano:1991ek,Sen:1991zi,Sen:1991cn, Meissner:1991zj,Meissner:1996sa,Codina:2021cxh}.
Based on these results, Hohm and Zwiebach conjectured that the standard
form of $O\left(d,d\right)$ transformations also preserves for all
orders in $\alpha^{\prime}$, enabling the classification of all $O\left(d,d\right)$
invariant $\alpha^{\prime}$ corrections. Since the corresponding
$O\left(d,d\right)$ invariant action involves only first-order derivatives,
it can be exactly solved. Consequently, it raises the question of
whether black hole singularities can be cured based on this new development.
Recent studies have successfully resolved the big bang singularity
\cite{Wang:2019kez,Wang:2019dcj,Wang:2019mwi,Gasperini:2023tus,Conzinu:2023fth,Bernardo:2019bkz}
and the curvature singularities of two-dimensional string black holes
\cite{Ying:2022xaj,Ying:2022cix,Codina:2023fhy,Codina:2023nwz} in
this framework. However, this approach still cannot resolve the singularity
of the Schwarzschild and rotating, accreting black holes. The challenge
arises from the fact that the spherical part of the black hole metric
does not exhibit $O\left(d,d\right)$ symmetry. Therefore, directly
applying the Hohm-Zwiebach action becomes impossible.

Encouragingly, the general structure near the black hole singularity,
namely Kasner universe, exhibits $O\left(d,d\right)$ symmetry, opening
avenues for understanding the black hole singularity in non-perturbative
string theory. To be specific, let us recall the tree-level string
effective action:

\begin{equation}
I_{\mathrm{String}}=\frac{1}{16\pi G_{D}}\int d^{D}x\sqrt{-g}e^{-2\phi}\left(R+4\left(\partial\phi\right)^{2}\right).
\end{equation}

\noindent where we adopt $16\pi G_{D}=1$, $\phi\left(x\right)$ denotes
a physical dilaton and we set Kalb-Ramond field $b_{\mu\nu}=0$ for
simplicity. Moreover, when $\phi$ is a constant, the action reduces
to the Einstein-Hilbert action and we will consider the stringy corrections
to this action in this letter. On the other hand, since the spacetime
dimension of bosonic string theory is fixed to be $26$, we can see
the manifold as a direct product of the black hole and a compact internal
space \cite{Callan:1988hs}. Due to the fact that the decoupling of
the black hole and internal space hold for all order of $\alpha^{\prime}$
corrections \cite{Myers:1987qx}, we only need to consider the black
hole part in the following calculations. To obtain the Kasner universe
near the black hole singularity, we first consider the $D=n+3$ dimensional
Schwarzschild-Tangherlini black hole \cite{Brandenberger:1993mw,Griffiths:2009dfa}:

\begin{equation}
ds^{2}=-\left(1-\left(\frac{r_{0}}{r}\right)^{n}\right)dt^{2}+\frac{dr^{2}}{1-\left(\frac{r_{0}}{r}\right)^{n}}+r^{2}d\Omega_{n+1}^{2},
\end{equation}

\noindent where $d\Omega_{n+1}^{2}$ denotes the metric of an $\left(n+1\right)$
dimensional sphere, and $r_{0}$ is an event horizon. Inside the event
horizon, the metric can be transformed into the Kasner metric \emph{near
the singularity} \cite{Kasner:1921zz}:

\begin{equation}
ds^{2}=-d\tau^{2}+\stackrel[i=1]{d}{\sum}\tau^{2\beta_{i}}dx_{i}^{2}.\label{eq:Kasner}
\end{equation}

\noindent The spacelike singularity is now located at $\tau=0$ which
corresponds to $r=0$ of Schwarzschild black hole. Moreover, in order
to satisfy the vacuum Einstein's equations, the Kasner exponents $\beta_{i}$
must satisfy the following constraints

\begin{equation}
\stackrel[i=1]{d}{\sum}\beta_{i}^{2}=1,\qquad\stackrel[i=1]{d}{\sum}\beta_{i}=1.\label{eq:Schwarzschild}
\end{equation}

\noindent For the four-dimensional Schwarzschild black hole, we have
$\left(\beta_{1},\beta_{2},\beta_{3}\right)=\left(-\frac{1}{3},\frac{2}{3},\frac{2}{3}\right)$.
One may wonder whether the initial spherical symmetry is broken in
the Kasner metric (\ref{eq:Kasner}). Near the singularity, the curvature
becomes large enough that the evolution of the black hole geometry
at different spatial points decouples. It means that the partial differential
Einstein's equations become ordinary differential equations (ODEs)
with respect to time \cite{Belinsky:1970ew}. Therefore, there is
no reason to assume the initial spherical symmetry is also maintained
in ODEs' solution. The BKL approximation (with the Kasner metric as
its solution) is sufficient to describe this region. It depicts an
asymmetrically spherical shell chaotically collapsing near the singularity.

Apart from the Schwarzschild black hole, the Kerr black hole is highly
interesting. However, the presence of an inner horizon in a rotating
black hole transforms the spacelike singularity into a timelike singularity.
Poisson and Israel studied the mass inflation instability at the inner
horizon, which leads to the formation of a null singularity at the
inner horizon \cite{Poisson:1990eh}. In subsequent work, Burko found
that if the radiation power law does not drop off quickly, a spacelike
singularity rather than a null singularity will be formed. In other
words, the inner horizon can be superseded by the BKL singularity
\cite{Burko:2002fv}. Hamilton then numerically studied the evolution
of the inner horizon of a rotating, accreting black hole, which implies
that the spacetime almost undergoes a BKL collapse \cite{Hamilton:2017qls}.
Subsequently, the analytic model of this collapse, called the inflationary
Kasner solution, was constructed, including two Kasner epochs: an
inflation epoch with Kasner exponents $\left(\beta_{1},\beta_{2},\beta_{3}\right)=\left(1,0,0\right)$
and a collapse epoch characterized by Kasner exponents $\left(\beta_{1},\beta_{2},\beta_{3}\right)=\left(-\frac{1}{3},\frac{2}{3},\frac{2}{3}\right)$
approaching the spacelike singularity, which approximates a Schwarzschild
geometry \cite{Mcmaken:2021isj}.

Since the Kasner universe only depends on $\tau$, the string action
in this background possesses the $O\left(d,d\right)$ symmetry. This
allows us to study the effects of string theory non-perturbatively
in near singularity region and examine how it may potentially resolve
the black hole singularities (the previous discussions on anisotropic
Hohm-Zwiebach action can be found in references \cite{Bernardo:2020nol,Nunez:2020hxx,Gasperini:2023tus,Bieniek:2023ubx}.
Although the spacelike singularity of isotropic string cosmology has
been resolved in many references \cite{Gasperini:1992em,Antoniadis:1993jc},
the anisotropic case is highly non-trivial and lacks any progress
on its exact solution \cite{Bieniek:2023ubx}). Thus, let us use the
following ansatz:

\begin{eqnarray}
ds^{2} & = & -d\tau^{2}+g_{ij}dx^{i}dx^{j},\nonumber \\
g_{ij} & = & \mathrm{diag}\left(a_{1}\left(\tau\right)^{2},a_{2}\left(\tau\right)^{2},\ldots,a_{i}\left(\tau\right)^{2}\right),
\end{eqnarray}

\noindent where $i=1,\ldots,d$. The corresponding anisotropic Hohm-Zwiebach
action is given by:

\begin{eqnarray}
I_{\mathrm{HZ}} & = & \int d^{D}x\sqrt{-g}e^{-2\phi}\left(R+4\left(\partial\phi\right)^{2}\right.\nonumber \\
 &  & \left.+\frac{1}{4}\alpha^{\prime}\left(R^{\mu\nu\rho\sigma}R_{\mu\nu\rho\sigma}+\cdots\right)+\alpha^{\prime2}\left(\cdots\right)+\cdots\right)\nonumber \\
 & = & \int d\tau e^{-\Phi}\left(-\dot{\Phi}^{2}-\frac{1}{8}\mathrm{Tr}\left(\dot{\mathcal{S}}^{2}\right)\right.\nonumber \\
 &  & +\alpha^{\prime}c_{2,0}\mathrm{Tr}\left(\dot{\mathcal{S}}^{4}\right)+\alpha^{\prime2}c_{3,0}\mathrm{Tr}\left(\dot{\mathcal{S}}^{6}\right)\nonumber \\
 &  & +\alpha^{\prime3}\left[c_{4,0}\mathrm{Tr}\left(\dot{\mathcal{S}}^{8}\right)+c_{4,1}\left(\mathrm{Tr}\left(\dot{\mathcal{S}}^{4}\right)\right)^{2}\right]\nonumber \\
 &  & \left.+\cdots\right),\label{eq:Odd action with alpha}
\end{eqnarray}

\noindent where $\dot{f}\left(\tau\right)\equiv\partial_{\tau}f\left(\tau\right)$.
Moreover, there is no arbitrariness for this action as a string effective
action due to the form of this action is constrained by the symmetry
and has removed ambiguous coefficients through the field redefinition.
On the other hand, it has been verified that at order $\alpha^{\prime3}$
of Type II string theory \cite{Codina:2020kvj} and order $\alpha^{\prime}$
of general torus compactifications \cite{Eloy:2019hnl}, they are
consistent with (\ref{eq:Odd action with alpha}) with specific coefficients
$c_{i,j}$. The string vacuum is also a solution of this action \cite{Hohm:2019ccp,Wang:2019mwi}.
For a bosonic case, $c_{1,0}=-\frac{1}{8}$, $c_{2,0}=\frac{1}{64}$,
$c_{3,0}=-\frac{1}{3.2^{7}}$, $c_{4,0}=\frac{1}{2^{12}}-\frac{3}{2^{12}}\zeta\left(3\right)$,
$c_{4,1}=\frac{1}{2^{16}}+\frac{1}{2^{12}}\zeta\left(3\right)$ and
$c_{k>4}$'s are unknown \cite{Codina:2021cxh}. This action systematically
provides all-order higher curvature corrections to Einstein\textquoteright s
gravity. For example, the leading order is the Einstein-Hilbert action
($\phi=\mathrm{constant}$), and the first order of $\alpha^{\prime}$
correction is the Gauss-Bonnet term. Moreover, $O\left(d,d\right)$
invariant dilaton is defined by $\Phi\equiv2\phi-\log\sqrt{-g}$.
The matrix $\mathcal{S}$ is defined as

\begin{equation}
\mathcal{S}\equiv\left(\begin{array}{cc}
0 & g_{ij}\\
g_{ij}^{-1} & 0
\end{array}\right),\quad\mathrm{Tr}\left(\dot{\mathcal{S}}^{2k}\right)=\left(-4\right)^{k}2\stackrel[i=1]{d}{\sum}H_{i}^{2k},
\end{equation}

\noindent where $H_{i}\equiv\frac{\dot{a}_{i}\left(\tau\right)}{a_{i}\left(\tau\right)}$
denotes the Hubble parameter. In addition, the $O\left(d,d\right)$
symmetry of the action (\ref{eq:Odd action with alpha}) is manifested
by the following transformations:

\begin{equation}
\Phi\rightarrow\Phi,\qquad a_{i}\rightarrow a_{i}^{-1},\qquad H_{i}\rightarrow-H_{i}.
\end{equation}

\noindent However, it is challenging to obtain the solutions of EOM
from the action (\ref{eq:Odd action with alpha}) due to the multi-trace
terms. As an example, let us consider the action at the order of $\alpha^{\prime3}$
which introduces the first multi-trace term into the action (\ref{eq:Odd action with alpha}).
The action can be written as follows:

\begin{equation}
I_{\mathrm{HZ}}^{\left(3\right)}=\alpha^{\prime3}\left[c_{4,0}\mathrm{Tr}\left(\dot{\mathcal{S}}^{8}\right)+c_{4,1}\left(\mathrm{Tr}\left(\dot{\mathcal{S}}^{4}\right)\right)^{2}\right].
\end{equation}

\noindent For the isotropic background:

\begin{eqnarray}
I_{\mathrm{HZ}}^{\left(3\right)} & = & \alpha^{\prime3}\left(c_{4,0}\left(-4\right)^{4}2dH^{8}+c_{4,1}\left(-4\right)^{4}4d^{2}H^{8}\right)\nonumber \\
 & = & \alpha^{\prime3}\left(\mathrm{constant}\times H^{8}\right),
\end{eqnarray}

\noindent all orders of $\alpha^{\prime}$ corrections can be straightforwardly
summed in a simple manner proportional to $H^{2k}$. And it is easy
to obtain the corresponding EOM. But, for the anisotropic case:

\begin{equation}
I_{\mathrm{HZ}}^{\left(3\right)}=\alpha^{\prime3}\left(-4\right)^{4}\left[2c_{4,0}\stackrel[i=1]{d}{\sum}H_{i}^{8}+4c_{4,1}\left(\stackrel[i=1]{d}{\sum}H_{i}^{4}\right)^{2}\right],
\end{equation}

\noindent which cannot be summed together and the effects of multi-trace
cannot be neglected.

The aim of this letter is to calculate the non-perturbative and non-singular
black hole solution of the action (\ref{eq:Odd action with alpha})
with all-order higher curvature corrections. To calculate its solution,
we first recall the EOM of the action (\ref{eq:Odd action with alpha}):

\begin{eqnarray}
\ddot{\Phi}+\frac{1}{2}\stackrel[i=1]{d}{\sum}H_{i}f_{i}\left(H_{i}\right) & = & 0,\nonumber \\
\left(\frac{d}{d\tau}-\dot{\Phi}\right)f_{i}\left(H_{i}\right) & = & 0,\nonumber \\
\dot{\Phi}^{2}+\stackrel[i=1]{d}{\sum}g_{i}\left(H_{i}\right) & = & 0,\label{eq:HZ EOM}
\end{eqnarray}

\noindent where 

\begin{eqnarray}
f_{i}\left(H_{i}\right) & = & -2H_{i}-2\alpha^{\prime}H_{i}^{3}-2\alpha^{\prime2}H_{i}^{5}\nonumber \\
 &  & -8\alpha^{\prime3}4^{4}\left[2c_{4,0}H_{i}^{7}+4c_{4,1}\left(\stackrel[j=1]{d}{\sum}H_{j}^{4}\right)H_{i}^{3}\right]+\cdots,\nonumber \\
g_{i}\left(H_{i}\right) & = & -H_{i}^{2}-\frac{3}{2}\alpha^{\prime}H_{i}^{4}-\frac{5}{3}\alpha^{\prime2}H_{i}^{6}\nonumber \\
 &  & -7\alpha^{\prime3}4^{4}\left[2c_{4,0}H_{i}^{8}+4c_{4,1}\left(\stackrel[j=1]{d}{\sum}H_{j}^{4}\right)H_{i}^{4}\right]+\cdots,\nonumber \\
\label{eq:EOM fh gh}
\end{eqnarray}

\noindent and there is no additional constraint $\frac{d}{dH_{i}}g_{i}\left(H_{i}\right)=H_{i}\frac{d}{dH_{i}}f_{i}\left(H_{i}\right)$
for the anisotropic case. These EOM are consistent with refs. \cite{Nunez:2020hxx,Bieniek:2023ubx}.
When $\alpha^{\prime}=0$, the tree-level EOM (\ref{eq:HZ EOM}) have
a solution

\begin{eqnarray}
\Phi\left(\tau\right) & = & -\log\left(\tau-\tau_{0}\right)+\Phi_{0},\nonumber \\
a_{i}\left(\tau\right) & = & C_{i}\left(\tau-\tau_{0}\right)^{\beta_{i}},\nonumber \\
H_{i}\left(\tau\right) & = & \frac{\beta_{i}}{\tau-\tau_{0}},\label{eq:tree-level solution}
\end{eqnarray}

\noindent where $\tau_{0}$, $C_{i}$ and $\Phi_{0}$ are arbitrary
constants. This result covers the solution of Einstein's gravity (\ref{eq:Kasner})
since the physical dilaton $\phi$ is a constant. The corresponding
Kretschmann scalar is

\begin{equation}
R^{\mu\nu\rho\sigma}R_{\mu\nu\rho\sigma}=\frac{4}{\left(\tau-\tau_{0}\right)^{4}}\left(\frac{1}{2}\stackrel[i=1]{d}{\sum}\beta_{i}^{4}-2\stackrel[i=1]{d}{\sum}\beta_{i}^{3}+\frac{3}{2}\right).
\end{equation}

\noindent The spacelike singularity is located at $\tau=\tau_{0}$.
Now, our aim is to remove this singularity using $\alpha^{\prime}$
corrections in string theory by the following strategy:
\begin{itemize}
\item Firstly, we aim to calculate the perturbative solution of the EOM
(\ref{eq:HZ EOM}) in the perturbative regime as $\alpha^{\prime}\rightarrow0$.
\item Then, in the non-perturbative regime for an arbitrary $\alpha^{\prime}$,
we need to find a non-singular solution that satisfies the EOM (\ref{eq:HZ EOM})
and covers arbitrary higher orders of perturbative solution with the
\emph{multi-trace terms} as $\alpha^{\prime}\rightarrow0$ (the detailed
calculations can be found in the Appendix).\emph{ }
\end{itemize}
To calculate the perturbation solution of the EOM (\ref{eq:HZ EOM}),
we introduce a new variable $\Omega$:
\begin{equation}
\Omega\equiv e^{-\Phi},\label{eq:pertur notation}
\end{equation}

\noindent where $\dot{\Omega}=-\dot{\Phi}\Omega$ and $\ddot{\Omega}=\left(-\ddot{\Phi}+\dot{\Phi}^{2}\right)\Omega$.
And the EOM (\ref{eq:HZ EOM}) become

\noindent 
\begin{eqnarray}
\ddot{\Omega}-\left(\stackrel[i=1]{d}{\sum}h_{i}\left(H_{i}\right)\right)\Omega & = & 0,\nonumber \\
\frac{d}{d\tau}\left(\Omega f_{i}\left(H_{i}\right)\right) & = & 0,\nonumber \\
\dot{\Omega}^{2}+\left(\stackrel[i=1]{d}{\sum}g_{i}\left(H_{i}\right)\right)\Omega^{2} & = & 0,\label{eq:reEOM}
\end{eqnarray}

\noindent where we define a new function

\begin{equation}
h_{i}\left(H_{i}\right)\equiv\frac{1}{2}H_{i}f_{i}\left(H_{i}\right)-g\left(H_{i}\right)=\alpha^{\prime}\frac{1}{2}H_{i}^{4}+\cdots,
\end{equation}

\noindent Next, we assume the perturbative solutions of the EOM (\ref{eq:reEOM})
to be:

\begin{eqnarray}
\Omega\left(\tau\right) & = & \Omega_{\left(0\right)}\left(\tau\right)+\alpha^{\prime}\Omega_{\left(1\right)}\left(\tau\right)+\alpha^{\prime2}\Omega_{\left(2\right)}\left(\tau\right)+\cdots,\nonumber \\
H_{i}\left(\tau\right) & = & H_{i\left(0\right)}\left(\tau\right)+\alpha^{\prime}H_{i\left(1\right)}\left(\tau\right)+\alpha^{\prime2}H_{i\left(2\right)}\left(\tau\right)+\cdots,\nonumber \\
\end{eqnarray}

\noindent where we denote $\Omega_{i}$ and $H_{i}$ as the $i$th
order of the perturbative solutions. Therefore, the perturbative solution
can be calculated order by order, which is

\begin{eqnarray}
H_{i}\left(\tau\right) & = & \frac{\beta_{i}}{\tau-\tau_{0}}-\frac{4\beta_{i}^{3}+\beta_{i}\stackrel[j=1]{d}{\sum}\beta_{j}^{4}}{4\left(\tau-\tau_{0}\right)^{3}}\alpha^{\prime}+\cdots,\nonumber \\
\Omega\left(\tau\right) & = & \gamma\left(\tau-\tau_{0}\right)+\frac{\gamma}{4\left(\tau-\tau_{0}\right)}\left(\stackrel[i=1]{d}{\sum}\beta_{i}^{4}\right)\alpha^{\prime}+\cdots,\nonumber \\
\label{eq:perturbed solution}
\end{eqnarray}

\noindent where $\gamma$ is an integration constant. The leading-order
of (\ref{eq:perturbed solution}) covers the geometry near the Schwarzschild
singularity (\ref{eq:Kasner}). Moreover, due to equation (\ref{eq:pertur notation}),
we also have:

\begin{equation}
\Phi\left(\tau\right)=-\log\left(\gamma\left(\tau-\tau_{0}\right)\right)-\frac{1}{4\left(\tau-\tau_{0}\right)^{2}}\left(\stackrel[i=1]{d}{\sum}\beta_{i}^{4}\right)\alpha^{\prime}+\cdots.\label{eq:perturbed solution dilaton}
\end{equation}

\noindent Based on the perturbative solutions (\ref{eq:perturbed solution})
and (\ref{eq:perturbed solution dilaton}), we now construct a non-perturbative
\footnote{Our statement that this solution is non-perturbative does not rely
on the convergence of the series. Rather, it emphasizes that the logarithmic
resummation captures contributions that cannot be represented by any
finite-order polynomial in $\alpha^{\prime}$. In this sense, the
solution extends into the non-perturbative regime, effectively incorporating
all orders of $\alpha^{\prime}$ corrections. } and non-singular solution that satisfies the EOM (\ref{eq:HZ EOM}).
Assuming the coefficients $c_{k\leq n}$ are known, the dilaton solution
takes the form

\begin{equation}
\Phi\left(\tau\right)=-\frac{1}{2N}\log\left(\stackrel[k=0]{N}{\sum}\lambda_{k}\sigma^{2k}\right),\qquad\sigma^{2}\equiv2d\frac{\tau^{2}}{\alpha^{\prime}},\label{eq:dilaton solution}
\end{equation}

\noindent where $N\geq n-1$ is an arbitrary integer corresponding
to the truncation order of the $\alpha^{\prime}$ expansion, and $\lambda_{0}>0$,
$\lambda_{n}>0$. The coefficients $\lambda_{N}$, $\lambda_{N-1}$,
... , $\lambda_{N-n+1}$ are determined by $c_{1}$, ... , $c_{n}$,
while the remaining parameters $\lambda_{0}$, $\lambda_{1}$, ...
, $\lambda_{N-n}$ are free and can be chosen to violate the singularity
condition (\ref{eq:singular condition-1}), which we will introduce
shortly. For example, consider the first two terms of the expansion
of (\ref{eq:dilaton solution}) in the perturbative regime $\tau\rightarrow\infty$
(or equivalently $\alpha^{\prime}\rightarrow0$):

\begin{eqnarray}
\Phi\left(\frac{\tau}{\alpha^{\prime}}\rightarrow\infty\right) & = & -\frac{1}{2}\log\left(2d\frac{\tau^{2}}{\alpha^{\prime}}\lambda_{N}^{1/N}\right)\nonumber \\
 &  & -\frac{1}{2N}\frac{\lambda_{N-1}}{\lambda_{N}}\frac{1}{2d}\frac{\alpha^{\prime}}{\tau^{2}}+\cdots.\label{eq:perturbed of dilaton solution}
\end{eqnarray}

\noindent As discussed earlier, the $N$-th order in the $\alpha^{\prime}$
expansion becomes dominant, and the coefficients $\lambda_{N}$ and
$\lambda_{N-1}$ can be fixed by matching to the perturbative coefficients
$c_{1}$, $c_{2}$. Comparing this expansion with the perturbative
solution (\ref{eq:perturbed solution dilaton}), we determine 

\begin{eqnarray}
\lambda_{N} & = & \gamma^{2N}\left(\frac{\alpha^{\prime}}{2d}\right)^{N},\nonumber \\
\lambda_{N-1} & = & dN\gamma^{2N}\left(\frac{\alpha^{\prime}}{2d}\right)^{N}\stackrel[j=1]{d}{\sum}\beta_{j}^{4},\nonumber \\
 & ...,
\end{eqnarray}

\noindent where we have set $\tau_{0}=0$ for simplicity, and used
$c_{1}=-1/8$, $c_{2}=1/64$. Higher-order coefficients are listed
in the Appendix B. The corresponding functions $f_{i}$, $g_{i}$
and $H_{i}$ are given by
\begin{eqnarray}
f_{i} & = & -\frac{2\beta_{i}}{\sqrt{\alpha^{\prime}}}e^{\Phi\left(\tau\right)},\nonumber \\
g_{i} & = & -\dot{\Phi}^{2}\stackrel[k=1]{\infty}{\sum}\frac{b_{k}}{1+\sigma^{2k-2}},\nonumber \\
H_{i} & = & \frac{\sqrt{\alpha^{\prime}}\ddot{\Phi}}{e^{\Phi\left(\tau\right)}\beta_{i}}\stackrel[k=1]{\infty}{\sum}\frac{b_{k}}{1+\sigma^{2k-2}}\nonumber \\
 &  & -\frac{\sqrt{2d}\dot{\Phi}}{e^{\Phi\left(\tau\right)}\beta_{i}}\stackrel[k=1]{\infty}{\sum}\frac{k\sigma^{2k-1}b_{k+1}}{\left(1+\sigma^{2k}\right)^{2}},\label{eq:NN solution}
\end{eqnarray}
where $b_{k}\left(\beta_{i}\right)$ is a constant and satisfies the
following constraint:
\begin{equation}
\stackrel[i=1]{d}{\sum}b_{1}\left(\beta_{i}\right)=2,\qquad\stackrel[i=1]{d}{\sum}b_{k}\left(\beta_{i}\right)=0,\qquad k>1.
\end{equation}

\noindent The detailed expressions for $b_{k}\left(\beta_{i}\right)$
can be found in the Appendix. This solution is non-perturbative because
it holds for any value of $\alpha^{\prime}$ and includes all-order
higher curvature corrections. Furthermore, the associated Kretschmann
scalar is given by
\begin{eqnarray}
R^{\mu\nu\rho\sigma}R_{\mu\nu\rho\sigma} & = & 2\left(\stackrel[i=1]{d}{\sum}H_{i}^{2}\right)^{2}+2\left(\stackrel[i=1]{d}{\sum}H_{i}^{4}\right)\nonumber \\
 &  & +8\left(\stackrel[i=1]{d}{\sum}H_{i}^{2}\dot{H}_{i}\right)+4\left(\stackrel[i=1]{d}{\sum}\dot{H}_{i}^{2}\right)\nonumber \\
 & = & \left(\stackrel[k=0]{N}{\sum}\lambda_{k}\sigma^{2k}\right)^{-\varrho}F\left(\tau\right),
\end{eqnarray}
where $\varrho\equiv8-\frac{2}{N},$ and $F\left(\tau\right)$ denotes
a regular function. The explicit form of $F\left(\tau\right)$ together
with a detailed discussion of its possible singular behavior are given
in Eq. (\ref{eq:KreScalar}) in Appendix B. So the singularities for
the spacetime and dilaton appear if and only if

\begin{equation}
\mathrm{Singular\;condition}:\qquad\stackrel[k=0]{N}{\sum}\lambda_{k}\sigma^{2k}=0,\label{eq:singular condition-1}
\end{equation}

\noindent has real roots. Specifically, to match the recently known
coefficients $c_{4,0}$ and $c_{4,1}$ for the 4th order of $\alpha^{\prime}$
correction, and set $d=3$, we obtain

\begin{eqnarray}
\stackrel[k=0]{N}{\sum}\lambda_{k}\sigma^{2k} & = & 2099520\sigma^{6}+1283040\sigma^{4}+13680\sigma^{2}\nonumber \\
 &  & +15336\zeta\left(3\right)+61187.
\end{eqnarray}

\noindent This equation has no real roots for $\sigma$, indicating
that the singularities of the Schwarzschild and rotating, accreting
black holes are eliminated in the non-perturbative solution (see Appendix
for details). Moreover, in this condition, both the string coupling
$g_{s}=\exp\left(\phi\right)$ and the curvature invariant $R^{\mu\nu\rho\sigma}R_{\mu\nu\rho\sigma}$
in the Einstein frame remain regular. The detailed analysis is provided
in Appendix D. Indeed, since our ansatz (\ref{eq:dilaton solution})
is constructed under the assumption that the coefficients $c_{k\leq n}$
are known, the parameter $\lambda_{k}$ is determined by the condition
$c_{n\leq k}$. Considering any order $k$, although $\lambda_{n\leq k}$
are fixed by the coefficients $c_{n\leq k}$, we always have freedom
to choose $\lambda_{n>k}$ which violate the singular condition (\ref{eq:singular condition-1}),
and obtaining the non-singular solutions. It is worth noting that
this ansatz does not represent the unique or complete solution but
rather provides a constructive framework to investigate how higher-order
$\alpha^{\prime}$ corrections in the $O\left(d,d\right)$-invariant
effective action may contribute to singularity resolution. Once the
full set of coefficients $c_{k}$ becomes available from string theory
(e.g., in heterotic or bosonic models), the solution would need to
be correspondingly modified. Moreover, It is straightforward to verify
that this solution (\ref{eq:NN solution}) covers the perturbative
solution (\ref{eq:perturbed solution}) as $\alpha^{\prime}\rightarrow0$.
Finally, it is worth noting that our results focus on the resolution
of the singularity. The key point is that the $\alpha^{\prime}$ corrections
become significant only when the curvature grows sufficiently large.
In the vicinity of the singularity, the inverse gradients of the background
fields become comparable to the fundamental string length, such that
$\alpha^{\prime}R\geq1$, $\alpha^{\prime}\left(\nabla\phi\right)^{2}\geq1$,
and so on. In this high-curvature regime, the contributions from all
orders of $\alpha^{\prime}$ corrections must be included. Our non-perturbative
and non-singular solution is precisely obtained in this region. In
the low-curvature regime, where $\alpha^{\prime}\rightarrow0$, our
non-perturbative solution can be expanded perturbatively and reproduces
the familiar Schwarzschild black hole of Einstein's gravity at leading
order. Furthermore, we would like to emphasize that there are several
indications that the $O\left(d,d\right)$ invariance may also be preserved
in the low-curvature black-hole region, although this point is often
overlooked in the literature. The Schwarzschild black hole (with vanishing
dilaton) is T-dual to the string black hole (with a nontrivial dilaton),
as observed in \cite{Tseytlin:1991wr,Ginsparg:1992af,Exirifard:2004ey}.
This duality has recently been successfully incorporated into the
$O\left(d,d\right)$-invariant framework at large $D$ in our previous
work \cite{Ying:2024wrv}. It is therefore possible to investigate
whether the $O\left(d,d\right)$ invariance can be preserved throughout
the entire spacetime and whether the transition between the near-singularity
(high-curvature) region and the asymptotic (low-curvature) black-hole
region is smooth.

In conclusion, it is believed that the singularity geometry of the
Schwarzschild black hole solution in Einstein\textquoteright s gravity
can be effectively described by the Kasner universe. Since the Schwarzschild
black hole is a solution of the string low-energy effective action
with a constant dilaton, this property allows us to consistently introduce
higher-order $\alpha^{\prime}$ corrections, which modify the near-singularity
geometry while preserving the Schwarzschild solution at leading order
in the low-curvature regime. More specifically, since the Kasner universe
only depends on $\tau$, we used the Hohm-Zwiebach action to study
the singularity problem of black holes. The results indicated that
the singularities of Schwarzschild and rotating, accreting black holes
may be resolved by considering the $\alpha^{\prime}$ corrections
of string theory if the singular condition (\ref{eq:singular condition-1})
is broken by the known coefficients $c_{k}$. Nevertheless, we can
still make the solution non-singular by adjusting $\lambda_{n>k}$
freely \cite{Wang:2019dcj}. Moreover, according to the BKL proposal,
the general realistic collapse to a spacelike singularity can be described
by the Kasner universe, which consists of different Kasner epochs
characterized by various choices of the Kasner exponents $\beta_{i}$
\cite{Belinsky:1970ew}. Since our results can apply to any $\beta_{i}$,
it suggests that it is possible to study the singularity problem for
the astrophysically realistic black holes through string theory. 

\vspace*{3.0ex}
\begin{acknowledgments}
\paragraph*{Acknowledgments.} 
We would especially like to thank Maurizio Gasperini, Robert Mann, Gabriele Veneziano for reading the draft and providing valuable suggestions and insights. We are also deeply indebt to Ronak Soni, Aron Wall, Peng Wang, Haitang Yang for many illuminating discussions and suggestions. HW is partially supported by NSFC Grant No.12275183, 12275184, 12105191 and 11875196. ZY is supported by an Internal Graduate Studentship of Trinity College, Cambridge and the AFOSR grant FA9550-19-1-0260 ``Tensor Networks and Holographic Spacetime''. SY is supported by NSFC Grant No.12105031 and cstc2021jcyj-bshX0227.
\end{acknowledgments}

\onecolumngrid
\newpage
\begin{center}
{\large{\bf Appendix}}
\end{center}

Here, we present the complete non-perturbative solution for the anisotropic
HZ action. Considering the following ansatz:

\begin{equation}
ds^{2}=-d\tau^{2}+\stackrel[i=1]{d}{\sum}a_{i}\left(\tau\right)^{2}dx_{i}^{2},
\end{equation}

\noindent the EOM for the anisotropic HZ action can be given as:

\begin{eqnarray}
\ddot{\Phi}+\frac{1}{2}\stackrel[i=1]{d}{\sum}H_{i}f_{i}\left(H_{i}\right) & = & 0,\nonumber \\
\left(\frac{d}{d\tau}-\dot{\Phi}^{2}\right)f_{i}\left(H_{i}\right) & = & 0,\nonumber \\
\dot{\Phi}^{2}+\stackrel[i=1]{d}{\sum}g_{i}\left(H_{i}\right) & = & 0,\label{eq:general EOM}
\end{eqnarray}

\subsection{\emph{Perturbative solutions up to an arbitrary higher order $\alpha^{\prime n}$:}}

\noindent The corresponding perturbative solution can be calculated
order by order, and it is given by:

\begin{eqnarray}
\Phi\left(\tau\right) & = & -\log\left(\gamma\tau\right)-\frac{\stackrel[j=1]{d}{\sum}\beta_{j}^{4}}{4}\frac{\alpha^{\prime}}{\tau^{2}}+\frac{9\left(\stackrel[j=1]{d}{\sum}\beta_{j}^{4}\right)^{2}+16\stackrel[j=1]{d}{\sum}\beta_{j}^{6}}{144}\frac{\alpha^{\prime2}}{\tau^{4}}\nonumber \\
 &  & -\frac{1}{720}\left[64\left(\stackrel[j=1]{d}{\sum}\beta_{j}^{4}\right)\left(\stackrel[k=1]{d}{\sum}\beta_{k}^{6}\right)+15\left(\stackrel[j=1]{d}{\sum}\beta_{j}^{4}\right)^{3}+72\stackrel[j=1]{d}{\sum}\beta_{j}^{8}\right.\nonumber \\
 &  & \left.+73728\left(\stackrel[k=1]{d}{\sum}\beta_{k}^{4}\right)^{2}c_{41}+36864\left(\stackrel[k=1]{d}{\sum}\beta_{k}^{8}\right)c_{40}\right]\frac{\alpha^{\prime3}}{\tau^{6}}+\cdots,\nonumber \\
H_{i}\left(\tau\right) & = & \frac{\beta_{i}}{\tau}-\frac{\beta_{i}\left(\stackrel[j=1]{d}{\sum}\beta_{j}^{4}+4\beta_{i}^{2}\right)}{4\tau^{3}}\alpha^{\prime}+\frac{\text{\ensuremath{\beta_{i}}}\left(576\beta_{i}^{4}+216\stackrel[j=1]{d}{\sum}\beta_{j}^{4}\beta_{i}^{2}+32\stackrel[j=1]{d}{\sum}\beta_{j}^{6}+27\left(\stackrel[j=1]{d}{\sum}\beta_{j}^{4}\right)^{2}\right)}{288\tau^{5}}\alpha^{\prime2}\nonumber \\
 &  & -\frac{\beta_{i}\alpha^{\prime3}}{1920\tau^{7}}\left[98304\left(\stackrel[j=1]{d}{\sum}\beta_{j}^{8}+40\beta_{i}^{6}\right)c_{4,0}+196608\stackrel[j=1]{d}{\sum}\beta_{j}^{4}\left(\stackrel[k=1]{d}{\sum}\beta_{k}^{4}+40\beta_{i}^{2}\right)c_{4,1}\right.\nonumber \\
 &  & +7680\beta_{i}^{6}+4800\left(\stackrel[j=1]{d}{\sum}\beta_{j}^{4}\right)\beta_{i}^{4}+640\left(\stackrel[j=1]{d}{\sum}\beta_{j}^{6}\right)\beta_{i}^{2}+900\left(\stackrel[j=1]{d}{\sum}\beta_{j}^{4}\right)^{2}\beta_{i}^{2}\nonumber \\
 &  & \left.+224\left(\stackrel[j=1]{d}{\sum}\beta_{j}^{4}\right)\left(\stackrel[k=1]{d}{\sum}\beta_{k}^{6}\right)+75\left(\stackrel[j=1]{d}{\sum}\beta_{j}^{4}\right)^{3}+192\stackrel[j=1]{d}{\sum}\beta_{j}^{8}\right]+\cdots,\label{eq:pur 3 order-1}
\end{eqnarray}

\noindent where $\gamma$ is an integration constant. Note that this
solution can be easily generalized to an arbitrary higher order using
Mathematica.

\subsection{\emph{Non-perturbative solutions which cover $\alpha^{\prime n}$
as $\alpha^{\prime}\rightarrow0$: }}

\noindent The non-perturbative solution of EOM (\ref{eq:general EOM})
is given by:

\noindent 
\begin{eqnarray}
\Phi\left(\tau\right) & = & -\frac{1}{2N}\log\left(\stackrel[k=0]{N}{\sum}\lambda_{k}\sigma^{2k}\right),\qquad\sigma^{2}\equiv2d\frac{\tau^{2}}{\alpha^{\prime}},\nonumber \\
f_{i}\left(H_{i}\right) & = & -\frac{2\beta_{i}}{\sqrt{\alpha^{\prime}}}\left(\stackrel[k=0]{N}{\sum}\lambda_{k}\sigma^{2k}\right)^{-\frac{1}{2N}},\nonumber \\
g_{i}\left(H_{i}\right) & = & -\frac{d}{2\alpha^{\prime}N^{2}}\left(\frac{\stackrel[k=1]{N}{\sum}2k\lambda_{k}\sigma^{2k-1}}{\stackrel[k=0]{N}{\sum}\lambda_{k}\sigma^{2k}}\right)^{2}\stackrel[k=1]{\infty}{\sum}\frac{b_{k}}{1+\sigma^{2k-2}},\nonumber \\
H_{i} & = & \frac{2d}{2N\beta_{i}\sqrt{\alpha^{\prime}}\left(\stackrel[k=0]{N}{\sum}\lambda_{k}\sigma^{2k}\right)^{2-\frac{1}{2N}}}\left[-\left(\stackrel[k=1]{N}{\sum}2k\left(2k-1\right)\lambda_{k}\sigma^{2k-2}\right)\left(\stackrel[k=0]{N}{\sum}\lambda_{k}\sigma^{2k}\right)\stackrel[k=1]{\infty}{\sum}\frac{b_{k}}{1+\sigma^{2k-2}}\right.\nonumber \\
 &  & \left.+\left(\stackrel[k=1]{N}{\sum}2k\lambda_{k}\sigma^{2k-1}\right)^{2}\stackrel[k=1]{\infty}{\sum}\frac{b_{k}}{1+\sigma^{2k-2}}+\left(\stackrel[k=1]{N}{\sum}2k\lambda_{k}\sigma^{2k-1}\right)\stackrel[k=0]{N}{\sum}\lambda_{k}\sigma^{2k}\stackrel[k=1]{\infty}{\sum}\frac{kb_{k+1}\sigma^{2k-1}}{\left(1+\sigma^{2k}\right)^{2}}\right].\label{eq:np solution}
\end{eqnarray}

\noindent In the perturbative region, corresponding to $\tau\rightarrow\infty$
(or equivalently $\alpha^{\prime}\rightarrow0$), the non-perturbative
solution can be expanded as follows:

\noindent 
\begin{eqnarray}
 &  & \Phi\left(\frac{\tau}{\alpha^{\prime}}\rightarrow\infty\right)\nonumber \\
 & = & -\frac{1}{2}\log\left(2d\frac{\tau^{2}}{\alpha^{\prime}}\lambda_{N}^{1/N}\right)-\frac{1}{2N}\nonumber \\
 &  & \times\left(\frac{\lambda_{N-1}}{\lambda_{N}}\frac{1}{2d}\frac{\alpha^{\prime}}{\tau^{2}}+\frac{2\lambda_{N}\lambda_{N-2}-\lambda_{N-1}^{2}}{2\lambda_{N}^{2}}\frac{1}{\left(2d\right)^{2}}\frac{\alpha^{\prime2}}{\tau^{4}}+\frac{3\lambda_{N}^{2}\lambda_{N-3}-3\lambda_{N}\lambda_{N-1}\lambda_{N-2}+\lambda_{N-1}^{3}}{3\lambda_{N}^{3}}\frac{1}{\left(2d\right)^{3}}\frac{\alpha^{\prime3}}{\tau^{6}}\right.\nonumber \\
 &  & \left.+\frac{4\lambda_{N}^{3}\lambda_{N-4}-2\lambda_{N}^{2}\lambda_{N-2}^{2}-4\lambda_{N}^{2}\lambda_{N-1}\lambda_{N-3}+4\lambda_{N}\lambda_{N-1}^{2}\lambda_{N-2}-\lambda_{N-1}^{4}}{4\lambda_{N}^{4}}\frac{1}{\left(2d\right)^{4}}\frac{\alpha^{\prime4}}{\tau^{8}}+\mathcal{O}\left(\frac{1}{\tau^{10}}\right)\right),
\end{eqnarray}

\noindent and

\noindent 
\begin{eqnarray}
 &  & H_{i}\left(\frac{\tau}{\alpha^{\prime}}\rightarrow\infty\right)\nonumber \\
 & = & \frac{b_{1}d^{\frac{1}{2}}\lambda_{N}^{\frac{1}{2N}}}{\sqrt{2}\beta_{i}}\frac{1}{\tau}+\frac{\left(8N\lambda_{N}b_{2}-5\lambda_{N-1}b_{1}\right)\left(d^{\frac{-1}{2}}\lambda_{N}^{\frac{1}{2N}-1}\right)}{4N\sqrt{2}\beta_{i}}\frac{\alpha^{\prime}}{\tau^{3}}\nonumber \\
 &  & -\frac{-\left(38N-11\right)\lambda_{N-1}^{2}b_{1}+48N\lambda_{N}\lambda_{N-1}b_{2}+48N^{2}\lambda_{N}^{2}b_{2}-48N^{2}\lambda_{N}^{2}b_{3}+76N\lambda_{N-2}\lambda_{N}b_{1}}{N^{2}2^{5}\sqrt{2}\beta_{i}}\left(d^{\frac{-3}{2}}\lambda_{N}^{\frac{1}{2N}-2}\right)\frac{\alpha^{\prime2}}{\tau^{5}}\nonumber \\
 &  & +\left[-\left(1262N^{2}-5133N+6560\right)\lambda_{N-1}^{3}b_{1}+4N^{2}\left(82N-25\right)\lambda_{N-2}\lambda_{N}\lambda_{N-1}b_{1}-328N^{3}\lambda_{N-3}\lambda_{N}^{2}b_{1}\right.\nonumber \\
 &  & +8N^{2}\left(22N-7\right)\lambda_{N}\lambda_{N-1}^{2}b_{2}+112N^{3}\lambda_{N}^{2}\lambda_{N-1}b_{2}+128N^{4}\lambda_{N}^{3}b_{2}-352N^{3}\lambda_{N-2}\lambda_{N}^{2}b_{2}\nonumber \\
 &  & \left.-112N^{3}\lambda_{N}^{2}\lambda_{N-1}b_{3}+128N^{4}\lambda_{N}^{3}b_{4}\right]\frac{d^{\frac{-5}{2}}\lambda_{N}^{\frac{1}{2N}-3}}{128N^{4}\sqrt{2}\beta_{i}}\frac{\alpha^{\prime3}}{\tau^{7}}+\mathcal{O}\left(\frac{1}{\tau^{9}}\right).
\end{eqnarray}
To compare this result with perturbative solution (\ref{eq:pur 3 order-1})
as $\alpha^{\prime}\rightarrow0$, we can fix $\gamma=\frac{1}{\sqrt{\alpha^{\prime}}}$
and the coefficients $\lambda_{k}$ and $b_{k}$: 
\begin{eqnarray}
\lambda_{N} & = & \left(\frac{1}{2d}\right)^{N},\nonumber \\
\lambda_{N-1} & = & dN\left(\frac{1}{2d}\right)^{N}\stackrel[j=1]{d}{\sum}\beta_{j}^{4},\nonumber \\
\lambda_{N-2} & = & \left(\frac{1}{2}\left(N-1\right)\left(\stackrel[j=1]{d}{\sum}\beta_{j}^{4}\right)^{2}-\frac{8}{9}\stackrel[j=1]{d}{\sum}\beta_{j}^{6}\right)d^{2}N\left(\frac{1}{2d}\right)^{N},\nonumber \\
\ensuremath{\lambda_{N-3}} & = & \left[\left(\frac{64}{45}-\frac{8}{9}N\right)\stackrel[j=1]{d}{\sum}\beta_{j}^{6}\stackrel[k=1]{d}{\sum}\beta_{k}^{4}+\frac{8}{5}\stackrel[j=1]{d}{\sum}\beta_{j}^{8}+\frac{N^{2}-3N+2}{6}\left(\stackrel[j=1]{d}{\sum}\beta_{j}^{4}\right)^{3}\right.\nonumber \\
 &  & \left.+\frac{8192}{5}\left(\stackrel[k=1]{d}{\sum}\beta_{k}^{4}\right)^{2}c_{4,1}+\frac{4096}{5}\left(\stackrel[k=1]{d}{\sum}\beta_{k}^{8}\right)c_{4,0}\right]d^{3}N\left(\frac{1}{2d}\right)^{N},\nonumber \\
\ldots
\end{eqnarray}

\noindent and

\noindent 
\begin{eqnarray}
b_{1}\left(\beta_{i}\right) & = & 2\beta_{i}^{2},\nonumber \\
b_{2}\left(\beta_{i}\right) & = & -d\beta_{i}^{2}\left(\beta_{i}^{2}-\stackrel[j=1]{d}{\sum}\beta_{j}^{4}\right),\nonumber \\
b_{3}\left(\beta_{i}\right) & = & \frac{d\beta_{i}^{2}}{3}\left(3\stackrel[j=1]{d}{\sum}\beta_{k}^{4}-8d\stackrel[j=1]{d}{\sum}\beta_{j}^{6}+8d\beta_{i}^{4}-3\beta_{i}^{2}\right),\nonumber \\
b_{4}\left(\beta_{i}\right) & = & \frac{d\beta_{i}^{2}}{9}\left[8d^{2}\left(\stackrel[j=1]{d}{\sum}\beta_{j}^{6}\right)\left(\stackrel[k=1]{d}{\sum}\beta_{k}^{4}\right)+72d^{2}\left(\stackrel[j=1]{d}{\sum}\beta_{j}^{8}\right)-9\left(\stackrel[j=1]{d}{\sum}\beta_{j}^{4}\right)\right.\nonumber \\
 &  & +\beta_{i}^{2}\left(16d^{2}\left(\stackrel[j=1]{d}{\sum}\beta_{j}^{6}\right)+9\right)-24d^{2}\beta_{i}^{4}\left(\stackrel[j=1]{d}{\sum}\beta_{j}^{4}\right)-72d^{2}\beta_{i}^{6}\nonumber \\
 &  & \left.+73728d^{2}c_{4,1}\left(\stackrel[j=1]{d}{\sum}\beta_{j}^{4}\right)\left(\left(\stackrel[k=1]{d}{\sum}\beta_{k}^{4}\right)-\beta_{i}^{2}\right)+36864d^{2}c_{4,0}\left(\stackrel[j=1]{d}{\sum}\beta_{j}^{8}-\beta_{i}^{6}\right)\right],\nonumber \\
\ldots,
\end{eqnarray}
where
\begin{equation}
\stackrel[i=1]{d}{\sum}b_{1}\left(\beta_{i}\right)=2,\qquad\stackrel[i=1]{d}{\sum}b_{k}\left(\beta_{i}\right)=0,\qquad k>1,
\end{equation}

\noindent Now, we study the singularity structure of the non-perturbative
solution (\ref{eq:np solution}) through the associated Kretschmann
scalar:

\noindent 
\begin{eqnarray}
R^{\mu\nu\rho\sigma}R_{\mu\nu\rho\sigma} & = & 2\left(\stackrel[i=1]{d}{\sum}H_{i}^{2}\right)^{2}+2\left(\stackrel[i=1]{d}{\sum}H_{i}^{4}\right)+8\left(\stackrel[i=1]{d}{\sum}H_{i}^{2}\dot{H}_{i}\right)+4\left(\stackrel[i=1]{d}{\sum}\dot{H}_{i}^{2}\right)\nonumber \\
 & = & 2\left[\left(\stackrel[i=1]{d}{\sum}\frac{1}{\beta_{i}^{2}}\right)^{2}+\stackrel[i=1]{d}{\sum}\frac{1}{\beta_{i}^{4}}\right]\frac{1}{e^{4\Phi\left(\tau\right)}}\left(\sqrt{\alpha^{\prime}}\ddot{\Phi}A\left(\sigma\right)-\sqrt{2d}\dot{\Phi}B\left(\sigma\right)\right)^{4}\nonumber \\
 &  & +8\stackrel[i=1]{d}{\sum}\frac{1}{\beta_{i}^{3}}\frac{1}{e^{3\Phi\left(\tau\right)}}\left(\sqrt{\alpha^{\prime}}\ddot{\Phi}A\left(\sigma\right)-\sqrt{2d}\dot{\Phi}B\left(\sigma\right)\right)^{2}\nonumber \\
 &  & \times\left[\sqrt{\alpha^{\prime}}\dddot{\Phi}A\left(\sigma\right)-\sqrt{\alpha^{\prime}}\ddot{\Phi}\dot{\Phi}A\left(\sigma\right)+\sqrt{\alpha^{\prime}}\ddot{\Phi}\frac{d}{d\tau}A\left(\sigma\right)\right.\nonumber \\
 &  & \left.-\sqrt{2d}\ddot{\Phi}B\left(\sigma\right)+\sqrt{2d}\dot{\Phi}^{2}B\left(\sigma\right)-\sqrt{2d}\dot{\Phi}\frac{d}{d\tau}B\left(\sigma\right)\right]\nonumber \\
 &  & +4\stackrel[i=1]{d}{\sum}\frac{1}{\beta_{i}^{2}}\frac{1}{e^{2\Phi\left(\tau\right)}}\left[\sqrt{\alpha^{\prime}}\dddot{\Phi}A\left(\sigma\right)-\sqrt{\alpha^{\prime}}\ddot{\Phi}\dot{\Phi}A\left(\sigma\right)+\sqrt{\alpha^{\prime}}\ddot{\Phi}\frac{d}{d\tau}A\left(\sigma\right)\right.\nonumber \\
 &  & \left.-\sqrt{2d}\ddot{\Phi}B\left(\sigma\right)+\sqrt{2d}\dot{\Phi}^{2}B\left(\sigma\right)-\sqrt{2d}\dot{\Phi}\frac{d}{d\tau}B\left(\sigma\right)\right]^{2}.
\end{eqnarray}
 For simplicity, we define
\begin{equation}
A\left(\sigma\right)\equiv\stackrel[k=1]{\infty}{\sum}\frac{b_{k}}{1+\sigma^{2k-2}},\;B\left(\sigma\right)\equiv\stackrel[k=1]{\infty}{\sum}\frac{k\sigma^{2k-1}b_{k+1}}{\left(1+\sigma^{2k}\right)^{2}}.\label{eq:HAB}
\end{equation}
To analyze the singularities of the Kretschmann scalar, we introduce
\begin{equation}
S\left(\sigma\right)\equiv\stackrel[k=0]{N}{\sum}\lambda_{k}\sigma^{2k},\label{eq:phiS}
\end{equation}
in terms of which the scalar curvature invariant takes the compact
form
\begin{equation}
R^{\mu\nu\rho\sigma}R_{\mu\nu\rho\sigma}=\left(\stackrel[k=0]{N}{\sum}\lambda_{k}\sigma^{2k}\right)^{-\varrho}F\left(\tau\right),\label{eq:RR}
\end{equation}
where $\varrho\equiv8-\frac{2}{N}$ and

\noindent 
\begin{eqnarray}
 &  & F\left(\tau\right)\nonumber \\
 & = & \frac{2}{\left(2N\right)^{4}}\left(\left(\stackrel[i=1]{d}{\sum}\frac{1}{\beta_{i}^{2}}\right)^{2}+\stackrel[i=1]{d}{\sum}\frac{1}{\beta_{i}^{4}}\right)\left(-\sqrt{\alpha^{\prime}}\left[S\left(\tau\right)\frac{d^{2}}{d\tau^{2}}S\left(\tau\right)-\left(\frac{d}{d\tau}S\left(\tau\right)\right)^{2}\right]A\left(\sigma\right)+\sqrt{2d}B\left(\sigma\right)S\left(\tau\right)\frac{d}{d\tau}S\left(\tau\right)\right)^{4}\nonumber \\
 &  & +8\stackrel[i=1]{d}{\sum}\frac{1}{\beta_{i}^{3}}\frac{1}{\left(2N\right)^{4}}S\left(\tau\right)^{1-\frac{1}{2N}}\left(-\sqrt{\alpha^{\prime}}\left[S\left(\tau\right)\frac{d^{2}}{d\tau^{2}}S\left(\tau\right)-\left(\frac{d}{d\tau}S\left(\tau\right)\right)^{2}\right]A\left(\sigma\right)+\sqrt{2d}B\left(\sigma\right)S\left(\tau\right)\frac{d}{d\tau}S\left(\tau\right)\right)^{2}\nonumber \\
 &  & \times\left[-\sqrt{\alpha^{\prime}}\left[2NS\left(\tau\right)^{2}\frac{d^{3}}{d\tau^{3}}S\left(\tau\right)-\left(6N-1\right)S\left(\tau\right)\frac{d}{d\tau}S\left(\tau\right)\frac{d^{2}}{d\tau^{2}}S\left(\tau\right)+\left(4N-1\right)\left(\frac{d}{d\tau}S\left(\tau\right)\right)^{3}\right]A\left(\sigma\right)\right.\nonumber \\
 &  & -2N\sqrt{\alpha^{\prime}}\left[S\left(\tau\right)^{2}\frac{d^{2}}{d\tau^{2}}S\left(\tau\right)-S\left(\tau\right)\left(\frac{d}{d\tau}S\left(\tau\right)\right)^{2}\right]\frac{d}{d\tau}A\left(\sigma\right)\nonumber \\
 &  & \left.+\sqrt{2d}\left[2NS\left(\tau\right)^{2}\frac{d^{2}}{d\tau^{2}}S\left(\tau\right)-\left(2N-1\right)S\left(\tau\right)\left(\frac{d}{d\tau}S\left(\tau\right)\right)^{2}\right]B\left(\sigma\right)+2N\sqrt{2d}S\left(\tau\right)^{2}\frac{d}{d\tau}S\left(\tau\right)\frac{d}{d\tau}B\left(\sigma\right)\right]\nonumber \\
 &  & +4\stackrel[i=1]{d}{\sum}\frac{1}{\beta_{i}^{2}}\frac{1}{\left(2N\right)^{4}}S\left(\tau\right)^{2-\frac{1}{N}}\nonumber \\
 &  & \times\left[-\sqrt{\alpha^{\prime}}\left[2NS\left(\tau\right)^{2}\frac{d^{3}}{d\tau^{3}}S\left(\tau\right)-\left(6N-1\right)S\left(\tau\right)\frac{d}{d\tau}S\left(\tau\right)\frac{d^{2}}{d\tau^{2}}S\left(\tau\right)+\left(4N-1\right)\left(\frac{d}{d\tau}S\left(\tau\right)\right)^{3}\right]A\left(\sigma\right)\right.\nonumber \\
 &  & -2N\sqrt{\alpha^{\prime}}\left[S\left(\tau\right)^{2}\frac{d^{2}}{d\tau^{2}}S\left(\tau\right)-S\left(\tau\right)\left(\frac{d}{d\tau}S\left(\tau\right)\right)^{2}\right]\frac{d}{d\tau}A\left(\sigma\right)\nonumber \\
 &  & \left.+\sqrt{2d}\left[2NS\left(\tau\right)^{2}\frac{d^{2}}{d\tau^{2}}S\left(\tau\right)-\left(2N-1\right)S\left(\tau\right)\left(\frac{d}{d\tau}S\left(\tau\right)\right)^{2}\right]B\left(\sigma\right)+2N\sqrt{2d}S\left(\tau\right)^{2}\frac{d}{d\tau}S\left(\tau\right)\frac{d}{d\tau}B\left(\sigma\right)\right]^{2}.\label{eq:KreScalar}
\end{eqnarray}
Using (\ref{eq:HAB}) and (\ref{eq:phiS}), one finds
\begin{eqnarray}
\frac{d}{d\tau}S\left(\sigma\right) & = & \stackrel[k=0]{N}{\sum}2k\lambda_{k}\sigma^{2k-1}\sqrt{\frac{2d}{\alpha^{\prime}}},\nonumber \\
\frac{d^{2}}{d\tau^{2}}S\left(\sigma\right) & = & \stackrel[k=0]{N}{\sum}2k\left(2k-1\right)\lambda_{k}\sigma^{2k-2}\frac{2d}{\alpha^{\prime}},\nonumber \\
\frac{d^{3}}{d\tau^{3}}S\left(\sigma\right) & = & \stackrel[k=0]{N}{\sum}2k\left(2k-1\right)\left(2k-2\right)\lambda_{k}\sigma^{2k-3}\left(\frac{2d}{\alpha^{\prime}}\right)^{\frac{3}{2}},\nonumber \\
\frac{d}{d\tau}A\left(\sigma\right) & = & -\stackrel[k=1]{\infty}{\sum}\frac{\left(2k-2\right)b_{k}\sigma^{2k-3}}{\left(1+\sigma^{2k-2}\right)^{2}},\nonumber \\
\frac{d}{d\tau}B\left(\sigma\right) & = & \stackrel[k=1]{\infty}{\sum}\frac{k\left(2k-1\right)\sigma^{2k-2}b_{k+1}\left(1+\sigma^{2k}\right)^{2}\sqrt{\frac{2d}{\alpha^{\prime}}}-\left(2k\right)^{2}\sigma^{2k-1}b_{k+1}\left(1+\sigma^{2k}\right)\sigma^{2k-1}\sqrt{\frac{2d}{\alpha^{\prime}}}}{\left(1+\sigma^{2k}\right)^{4}}.\nonumber \\
\label{eq:dSAB}
\end{eqnarray}

\noindent It is evident that (\ref{eq:HAB}), (\ref{eq:phiS}) and
(\ref{eq:dSAB}) are free from singularities, since the presence of
$1+\sigma^{2n}$ (with even powers of $\sigma$) in the denominator
never vanishes. Therefore, $F\left(\tau\right)$, which is constructed
from finite combinations of these quantities through addition, subtraction,
and multiplication, is a regular function. Consequently, the spacetime
and dilaton fields become singular if and only if

\begin{equation}
S\left(\sigma\right)=\stackrel[k=0]{N}{\sum}\lambda_{k}\sigma^{2k}=0,\label{eq:singular condition}
\end{equation}

\noindent admits real roots.

\subsection{\emph{Non-perturbative solutions which cover $\alpha^{\prime3}$
as $\alpha^{\prime}\rightarrow0$ (special example): }}

To cover the perturbative solution (\ref{eq:pur 3 order-1}) up to
the order of $\alpha^{\prime3}$ as $\alpha^{\prime}\rightarrow0$,
and including the known coefficients $c_{1,0}=-\frac{1}{8}$, $c_{2,0}=\frac{1}{64}$,
$c_{3,0}=-\frac{1}{3.2^{7}}$, $c_{4,0}=\frac{1}{2^{12}}-\frac{3}{2^{12}}\zeta\left(3\right)$,
$c_{4,1}=\frac{1}{2^{16}}+\frac{1}{2^{12}}\zeta\left(3\right)$, we
need to choose $N=3$ and $d=3$, which denotes the singularity regions
of the four-dimensional Schwarzschild and rotating, accreting black
holes. The corresponding non-perturbative solution (\ref{eq:np solution})
is given by:

\begin{eqnarray}
\Phi\left(\tau\right) & = & -\frac{1}{6}\log\left(\lambda_{0}+\lambda_{1}\sigma^{2}+\lambda_{2}\sigma^{4}+\lambda_{3}\sigma^{6}\right),\qquad\sigma^{2}\equiv6\frac{\tau^{2}}{\alpha^{\prime}},\nonumber \\
f_{i}\left(H_{i}\right) & = & -\frac{2\beta_{i}}{\sqrt{\alpha^{\prime}}}\left(\lambda_{0}+\lambda_{1}\sigma^{2}+\lambda_{2}\sigma^{4}+\lambda_{3}\sigma^{6}\right)^{-\frac{1}{6}},\nonumber \\
g_{i}\left(H_{i}\right) & = & -\frac{\left(\lambda_{1}\sigma+2\lambda_{2}\sigma^{3}+3\lambda_{3}\sigma^{5}\right)^{2}\left(\frac{b_{1}}{2}+\frac{b_{2}}{1+\sigma^{2}}+\frac{b_{3}}{1+\sigma^{4}}+\frac{b_{4}}{1+\sigma^{6}}\right)}{3\alpha^{\prime}\left(\lambda_{0}+\lambda_{1}\sigma^{2}+\lambda_{2}\sigma^{4}+\lambda_{3}\sigma^{6}\right)^{2}},\nonumber \\
H_{i}\left(\tau\right) & = & \frac{1}{\sqrt{\alpha^{\prime}}\beta_{i}}\left(\lambda_{0}+\lambda_{1}\sigma^{2}+\lambda_{2}\sigma^{4}+\lambda_{3}\sigma^{6}\right)^{-\frac{11}{6}}\left[-\left(\lambda_{0}+\lambda_{1}\sigma^{2}+\lambda_{2}\sigma^{4}+\lambda_{3}\sigma^{6}\right)\times\right.\nonumber \\
{\color{orange}} & {\color{orange}} & \left(2\lambda_{1}+12\lambda_{2}\sigma^{2}+30\lambda_{3}\sigma^{3}\right)\left(\frac{b_{1}}{2}+\frac{b_{2}}{1+\sigma^{2}}+\frac{b_{3}}{1+\sigma^{4}}+\frac{b_{4}}{1+\sigma^{6}}\right)\nonumber \\
{\color{orange}} & {\color{orange}} & +\left(2\lambda_{1}\sigma+4\lambda_{2}\sigma^{3}+6\lambda_{3}\sigma^{5}\right)\left(\lambda_{0}+\lambda_{1}\sigma^{2}+\lambda_{2}\sigma^{4}+\lambda_{3}\sigma^{6}\right)\times\nonumber \\
{\color{orange}} & {\color{orange}} & \left(\frac{\sigma b_{2}}{\left(1+\sigma^{2}\right)^{2}}+\frac{2\sigma^{3}b_{3}}{\left(1+\sigma^{4}\right)^{2}}+\frac{3\sigma^{5}b_{4}}{\left(1+\sigma^{6}\right)^{2}}\right)+\left(2\lambda_{1}\sigma+4\lambda_{2}\sigma^{3}+6\lambda_{3}\sigma^{5}\right)^{2}\times\nonumber \\
 &  & \left.\left(\frac{b_{1}}{2}+\frac{b_{2}}{1+\sigma^{2}}+\frac{b_{3}}{1+\sigma^{4}}+\frac{b_{4}}{1+\sigma^{6}}\right)\right],\label{eq:cover 3}
\end{eqnarray}

\noindent with $\lambda_{3}=\frac{1}{216}$, $\lambda_{2}=\frac{1}{216}$,
$\lambda_{1}=\frac{19}{17496}$, $\text{\ensuremath{\lambda_{0}}}=\frac{6303744c_{4,0}+26763264c_{4,1}+5701}{262440}$
and

\begin{eqnarray}
b_{1}\left(\beta_{i}\right) & = & 2\beta_{i}^{2},\nonumber \\
b_{2}\left(\beta_{i}\right) & = & 3\beta_{i}^{2}\left(\frac{11}{27}-\beta_{i}^{2}\right),\nonumber \\
b_{3}\left(\beta_{i}\right) & = & \beta_{i}^{2}\left(24\beta_{i}^{4}-3\beta_{i}^{2}-\frac{245}{81}\right),\nonumber \\
b_{4}\left(\beta_{i}\right) & = & \frac{\beta_{i}^{2}}{3}\left(270336c_{41}\left(\frac{11}{27}-\beta_{i}^{2}\right)+331776c_{40}\left(\frac{19}{243}-\beta_{i}^{6}\right)-648\beta_{i}^{6}-88\beta_{i}^{4}+\frac{931\beta_{i}^{2}}{27}+\frac{38047}{729}\right),
\end{eqnarray}

\noindent where $c_{4,0}=\frac{1}{2^{12}}-\frac{3}{2^{12}}\zeta\left(3\right)$
and $c_{4,1}=\frac{1}{2^{16}}+\frac{1}{2^{12}}\zeta\left(3\right)$.
It is worth noting that

\begin{equation}
\stackrel[i=1]{3}{\sum}b_{1}\left(\beta_{i}\right)=2,
\end{equation}
and
\begin{equation}
\stackrel[i=1]{3}{\sum}b_{2}\left(\beta_{i}\right)=\stackrel[i=1]{3}{\sum}b_{3}\left(\beta_{i}\right)=\stackrel[i=1]{3}{\sum}b_{4}\left(\beta_{i}\right)=0.
\end{equation}

\noindent The corresponding Kretschmann scalar is:
\begin{eqnarray}
R^{\mu\nu\rho\sigma}R_{\mu\nu\rho\sigma} & = & 2\left(\stackrel[i=1]{d}{\sum}H_{i}^{2}\right)^{2}+2\left(\stackrel[i=1]{d}{\sum}H_{i}^{4}\right)+8\left(\stackrel[i=1]{d}{\sum}H_{i}^{2}\dot{H}_{i}\right)+4\left(\stackrel[i=1]{d}{\sum}\dot{H}_{i}^{2}\right)\nonumber \\
 & = & \left(\left(15336\zeta\left(3\right)+61187\right)+13680\sigma^{2}+1283040\sigma^{4}+2099520\sigma^{6}\right)^{-\varrho}F\left(\tau\right),
\end{eqnarray}
Then, it is easy to check that this equation has no real root, implying
the singularities of four-dimensional Schwarzschild and rotating,
accreting black holes are removed. Finally, we wish to note that although
the coefficients $c_{k\geq5}$ are unknown, in principle, they can
be determined perturbatively by the worldsheet anomaly cancellation.

\subsection{Regularity of the string coupling and curvature invariant in the
Einstein frame}

To examine the regularity of the string coupling $g_{s}=\exp\left(\phi\right)$
, we first recall the definition of the physical dilaton,

\begin{eqnarray}
2\phi & = & \Phi+\log\sqrt{\left|g_{ij}\right|}\nonumber \\
 & = & -\frac{1}{2N}\log\stackrel[k=0]{N}{\sum}\lambda_{k}\sigma^{2k}+\stackrel[i=1]{d}{\sum}\int H_{i}d\tau,\label{eq:physical dilaton}
\end{eqnarray}

\noindent where the Hubble parameter is given by

\begin{equation}
H_{i}=\frac{S\left(\tau\right)^{\frac{1}{2N}-2}}{2N\beta_{i}}\left(-\sqrt{\alpha^{\prime}}A\left(\sigma\right)\left(S\left(\tau\right)\frac{d^{2}}{d\tau^{2}}S\left(\tau\right)-\left(\frac{d}{d\tau}S\left(\tau\right)\right)^{2}\right)+\sqrt{2d}B\left(\sigma\right)S\left(\tau\right)\frac{d}{d\tau}S\left(\tau\right)\right),
\end{equation}

\noindent with $A\left(\sigma\right)$ and $B\left(\sigma\right)$
defined in (\ref{eq:HAB}), and $S\left(\tau\right)$ given by (\ref{eq:phiS}).
To verify the regularity of the physical dilaton, we need to check
whether

\begin{equation}
S\left(\tau\right)=\stackrel[k=0]{N}{\sum}\lambda_{k}\sigma^{2k},
\end{equation}

\noindent vanishes. If $S\left(\tau\right)\neq0$, then the first
term in the expression (\ref{eq:physical dilaton}) for $\phi$ remains
regular. Furthermore, from the coordinate transformation between the
inner horizon of the Schwarzschild black hole and the Kasner universe,
the region where the Kasner description is valid is bounded by 

\begin{equation}
0<\tau\ll\tau_{max}.
\end{equation}

\noindent For instance, in the four-dimensional case, the maximal
value of $\tau$ is $\tau_{max}=\frac{4}{3}GM$. In this integration
domain, the integrand $H_{i}$ of $\int H_{i}d\tau$ is both continuous
and bounded, implying that the integral is convergent. Consequently,
the condition $S\left(\sigma\right)=\stackrel[k=0]{N}{\sum}\lambda_{k}\sigma^{2k}\neq0$
guarantees that $\phi$ is regular, and hence the string coupling
$g_{s}$ remains finite. Therefore, loop corrections need not be considered
in our analysis.

On the other hand, to study the regularity of the curvature invariant
$R^{\mu\nu\rho\sigma}R_{\mu\nu\rho\sigma}$ in the Einstein frame,
we recall the relation between the string frame and Einstein frame
metrics:

\begin{equation}
g_{\mu\nu}^{E}\LyXZeroWidthSpace=e^{-\frac{4\LyXZeroWidthSpace\left(\phi-\phi_{0}\right)\LyXZeroWidthSpace}{d-1}}g_{\mu\nu}^{\LyXZeroWidthSpace}.
\end{equation}

\noindent The corresponding Hubble parameter in the Einstein frame
is related to that in the string frame by

\begin{equation}
H_{i}^{E}=e^{\frac{2\phi}{d-1}}\left(H_{i}-\frac{2\dot{\phi}}{d-1}\right)
\end{equation}

\noindent Based on this relation, the curvature invariant in the Einstein
frame can be written as

\begin{eqnarray}
 &  & \left(R^{\mu\nu\rho\sigma}R_{\mu\nu\rho\sigma}\right)^{E}\nonumber \\
 & = & 2\left(\stackrel[i=1]{d}{\sum}\left(H_{i}^{E}\right)^{2}\right)^{2}+2\left(\stackrel[i=1]{d}{\sum}\left(H_{i}^{E}\right)^{4}\right)+8\left(\stackrel[i=1]{d}{\sum}\left(H_{i}^{E}\right)^{2}\frac{d}{dt}H_{i}^{E}\right)+4\left(\stackrel[i=1]{d}{\sum}\left(\frac{d}{dt}H_{i}^{E}\right)^{2}\right)\nonumber \\
 & = & 2e^{\frac{8\phi}{d-1}}\left(\stackrel[i=1]{d}{\sum}\left(H_{i}-\frac{1}{d-1}\left(\dot{\Phi}+\stackrel[j=1]{d}{\sum}H_{j}\right)\right)^{2}\right)^{2}+2e^{\frac{8\phi}{d-1}}\left(\stackrel[i=1]{d}{\sum}\left(H_{i}-\frac{1}{d-1}\left(\dot{\Phi}+\stackrel[j=1]{d}{\sum}H_{j}\right)\right)^{4}\right)\nonumber \\
 &  & +8e^{\frac{8\phi}{d-1}}\stackrel[i=1]{d}{\sum}\left(H_{i}-\frac{1}{d-1}\left(\dot{\Phi}+\stackrel[j=1]{d}{\sum}H_{j}\right)\right)^{2}\left[\frac{H_{i}}{d-1}\left(\dot{\Phi}+\stackrel[j=1]{d}{\sum}H_{j}\right)\right.\nonumber \\
 &  & \left.-\frac{1}{\left(d-1\right)^{2}}\left(\dot{\Phi}+\stackrel[j=1]{d}{\sum}H_{j}\right)^{2}+\dot{H}_{i}-\frac{1}{d-1}\left(\ddot{\Phi}+\stackrel[j=1]{d}{\sum}\dot{H}_{j}\right)\right]\nonumber \\
 &  & +4e^{\frac{8\phi}{d-1}}\stackrel[i=1]{d}{\sum}\left(\frac{H_{i}}{d-1}\left(\dot{\Phi}+\stackrel[j=1]{d}{\sum}H_{j}\right)-\frac{1}{\left(d-1\right)^{2}}\left(\dot{\Phi}+\stackrel[j=1]{d}{\sum}H_{j}\right)^{2}+\dot{H}_{i}-\frac{1}{d-1}\left(\ddot{\Phi}+\stackrel[j=1]{d}{\sum}\dot{H}_{j}\right)\right)^{2}.
\end{eqnarray}

\noindent The quantities involved are

\begin{eqnarray}
H_{i} & = & \frac{S\left(\tau\right)^{\frac{1}{2N}}}{\beta_{i}}\left(\sqrt{\alpha^{\prime}}\ddot{\Phi}A\left(\sigma\right)-\sqrt{2d}\dot{\Phi}B\left(\sigma\right)\right),\nonumber \\
\dot{H}_{i} & = & S\left(\tau\right)^{\frac{1}{2N}}\left(\frac{\sqrt{\alpha^{\prime}}\dddot{\Phi}}{\beta_{i}}A\left(\sigma\right)-\frac{\sqrt{\alpha^{\prime}}\ddot{\Phi}\dot{\Phi}}{\beta_{i}}A\left(\sigma\right)+\frac{\sqrt{\alpha^{\prime}}\ddot{\Phi}}{\beta_{i}}\frac{d}{d\tau}A\left(\sigma\right)\right.\nonumber \\
 &  & \left.-\frac{\sqrt{2d}\ddot{\Phi}}{\beta_{i}}B\left(\sigma\right)+\frac{\sqrt{2d}\dot{\Phi}^{2}}{\beta_{i}}B\left(\sigma\right)-\frac{\sqrt{2d}\dot{\Phi}}{\beta_{i}}\frac{d}{d\tau}B\left(\sigma\right)\right),
\end{eqnarray}

\noindent and

\noindent 
\begin{eqnarray}
\dot{\Phi} & = & -\frac{\frac{d}{d\tau}S\left(\tau\right)}{2NS\left(\tau\right)},\nonumber \\
\ddot{\Phi} & = & -\frac{S\left(\tau\right)\frac{d^{2}}{d\tau^{2}}S\left(\tau\right)-\left(\frac{d}{d\tau}S\left(\tau\right)\right)^{2}}{2NS\left(\tau\right)^{2}},\nonumber \\
\dddot{\Phi} & = & -\frac{S\left(\tau\right)^{2}\frac{d^{3}}{d\tau^{3}}S\left(\tau\right)-3S\left(\tau\right)\frac{d}{d\tau}S\left(\tau\right)\frac{d^{2}}{d\tau^{2}}S\left(\tau\right)+2\left(\frac{d}{d\tau}S\left(\tau\right)\right)^{3}}{2NS\left(\tau\right)^{3}}.
\end{eqnarray}

\noindent Following the same analysis as in Appendix B, we find that
the regularity of all quantities depends on whether $S\left(\tau\right)$
vanishes. If $S\left(\tau\right)=\stackrel[k=0]{N}{\sum}\lambda_{k}\sigma^{2k}\neq0$,
then all terms $\phi$, $H_{i},\dot{H}_{i},\dot{\Phi},\ddot{\Phi}$
and $\dddot{\Phi}$ are regular. Consequently, the curvature invariant
$\left(R^{\mu\nu\rho\sigma}R_{\mu\nu\rho\sigma}\right)^{E}$ is also
free of singularities in the Einstein frame.
\end{document}